\documentclass[amssymb,amsmath,prb,twocolumn,superscriptaddress]{revtex4}
\usepackage{graphicx}

\begin{document}

\title{Quantum simulation of Anderson and Kondo lattices with
  superconducting qubits}

\author{Juan Jos\'e Garc{\'\i}a-Ripoll}
\address{Universidad Complutense, Facultad de F{\'\i}sicas, Ciudad
  Universitaria s/n, Madrid, E-20808, Spain}

\author{Enrique Solano}
\address{Physics Department, CeNS and ASC,
Ludwig-Maximilians-Universit\"at, Theresienstrasse 37, 80333
Munich, Germany}

\address{Secci\'{o}n F\'{\i}sica, Departamento de Ciencias,
Pontificia Universidad Cat\'{o}lica del Per\'{u}, Apartado 1761,
Lima, Peru}

\author{Miguel Angel Martin-Delgado}
\address{Universidad Complutense, Facultad de F{\'\i}sicas, Ciudad
  Universitaria s/n, Madrid, E-20808, Spain}

\begin{abstract}
  We introduce a mapping between a variety of superconducting circuits
  and a family of Hamiltonians describing localized magnetic
  impurities interacting with conduction bands. This includes the Anderson model, the
  single impurity one- and two-channel Kondo problem, as well as the 1D Kondo lattice. We compare the
  requirements for performing quantum simulations using the proposed
  circuits to those of universal quantum computation with
  superconducting qubits, singling out the specific challenges that
  will have to be addressed.
\end{abstract}

\maketitle

\section{Introduction}

Quantum simulation~\cite{feynman82} consists on tuning the dynamics of
a flexible quantum mechanical system to simulate the properties of
another physical system or of a quantum mechanical model whose
solution is unknown. While a universal quantum computer can
efficiently simulate the dynamics of any quantum system
\cite{lloyd96,abrams97}, designing a quantum simulator for a
specific model should simplify the experimental requirements. In
particular, quantum simulators require neither high fidelity thresholds
nor error correction, and thus fewer qubits are needed to get
interesting results which are out of reach for classical numerical
computations. The successful simulation of the Bose-Hubbard model
using cold atoms in optical lattices~\cite{jaksch98,greiner02} is a
paradigmatic example.

In the last years, there have been tremendous experimental
achievements in the context of superconducting qubits, improving both
quantum control and coherence
times~\cite{bouchiat98,nakamura99,vion02,martinis02,chiorescu03,yamamoto03,pashkin03,duty04,wallraff04,fowler07}.
While the aim is the implementation of a scalable universal quantum
computer \cite{wallraff04,fowler07,helmer07}, this technology can be
used in the quantum simulation of many-body physics with
Josephson-Junction arrays \cite{oudenaarden96,fazio01,bruder05}.

With this motivation, we first design a circuit to simulate the
Anderson model for an impurity ion coupled to conduction electrons
\cite{anderson61}. This is done through an exact mapping from the
effective models describing the superconducting circuits to the
fermionic Hamiltonian for the impurity problem. In a certain parameter
regime of our circuit, this leads to the Kondo
Hamiltonian modeling the non-trivial physics of itinerant electrons that interact with a localized impurity having non-zero magnetic
moment~\cite{hewson97}. This is one of the central models for strongly
correlated electrons with implications going beyond Condensed Matter
Physics.  Furthermore, our constructions are versatile enough to
simulate other relevant Hamiltonians. With an appropriate choice of
the geometry, these include the Kondo lattice \cite{tsunetsugu97} and
the multichannel Kondo model \cite{nozieres80}. The latter exhibits
the simplest example of non-Fermi liquid behavior, which is a strong
deviation from the standard model of condensed matter systems
introduced by Landau.

We will also discuss the experimental challenges for implementing such
circuits. In particular, while coherence and interaction times of
current superconducting charge qubits seem good enough to simulate the
long-time dynamical and static properties of these lattices, further
experimental work is needed. Finally, we propose a set of measurements
to gather information about transport, correlation and energy spectra
and comment possible implications of the present work.

\section{Circuit design}

\subsection{Elementary components}

\begin{figure}
  \centering
  \includegraphics[width=0.95\linewidth]{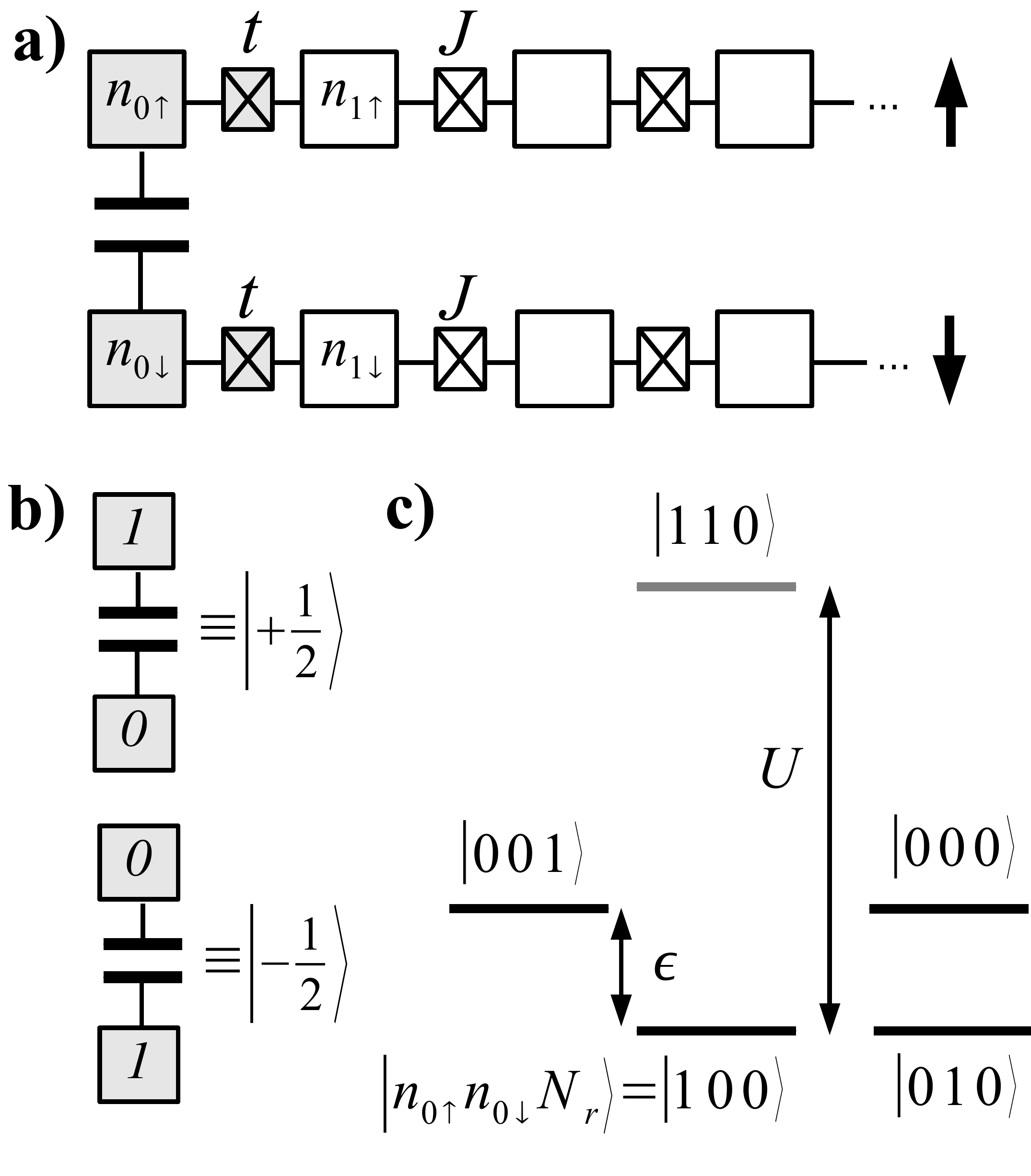}
  \caption{ (a) Quantum circuit simulating the single Anderson or
    Kondo impurity model. The big square boxes are superconducting
    islands, connected by crossed squares which are Josephson
    junctions. A selected pair of islands is capacitively coupled and
    represents a magnetic impurity. The Josephson energy for tunneling
    between junctions are $J$ for the main lattice and $t$ around the
    impurity. (b) Two occupation states for the impurity representing
    a pseudo-spin state. (c) Energy levels of the impurity and a
    neighboring site. $n_{0\uparrow}, n_{0\downarrow}$ and $N_r$
    represent the number of pairs of the two islands in the impurity
    and of the rest of the circuit. By external voltages,
    $-\epsilon/2$, and the capacitive coupling, $U\sim 1/C$, it
    becomes favorable to have a single pair in the impurity.}
  \label{fig:circuit}
\end{figure}

Let us start by describing the proposed architecture. We consider two
or more lines of low-capacitance superconducting islands coupled by
Josephson junctions [Fig.~\ref{fig:circuit}a]. While not shown here,
each island is itself part of a circuit like that of a charge qubit,
with an external lead which capacitively induces an offset potential
on the island, and possibly a Josephson coupling to a superconducting
reservoir of Cooper pairs\cite{blais04}.

Our quantum simulation protocols are based on two building blocks.
First, as we will show below, each island will have a very low
capacitance and be constrained to have at most one Cooper pair, which
can be treated as an impenetrable boson. Second, selected pairs of
islands will be coupled capacitively [Fig.~\ref{fig:circuit}b]. The
coupling and voltages of these islands will be adjusted to create a
suitable energy landscape that favors having a single excess Cooper
pair in either the upper or lower island, states which we associate
with an effective pseudo spin. The appropriate choice of energies has
the following form \begin{equation}
  E(n_{0\uparrow},n_{0\downarrow}) =
  U n_{0\uparrow} n_{0\downarrow} - \epsilon (n_{0\uparrow} + n_{0\downarrow}).
\end{equation}
Here, $n_{0\uparrow}$ and $n_{0\downarrow}$ represent the excess of
Cooper pairs on the upper and lower island in the impurity
[Fig.~\ref{fig:circuit}a-b]. The constants $U$ and $\epsilon$ denote,
respectively, the repulsive interaction energy coming from the
capacitive coupling between islands and an energy offset of these
islands with respect to the rest of the circuit.

As mentioned before, when we embed such an element in a circuit, the
energy offset $\epsilon$ enforces the impurity to host at most a
single excess Cooper pair. In Fig.~\ref{fig:circuit}c we have depicted
the lowest energy levels of a circuit containing a single impurity,
with occupation numbers $n_{0\uparrow}$ and $n_{0\downarrow}$,
connected to an array of ordinary islands, with total population
$N_r.$ A possible configuration is shown in
Fig.~\ref{fig:circuit}a. If we neglect the kinetic energy of hopping
pairs, we find that the states where the impurity is occupied by a
single pair, $n_{0\uparrow}=1$ or $n_{0\downarrow}=1$, have lower
energy than states where this pair is in the \textit{rest} of the
circuit. Multiply occupied impurities, $n_{0\uparrow}+n_{0\downarrow}
> 1$, have a even larger energy, $U$, so that we can neglect these
states.

\subsection{Quantum circuit model}
\label{sec:qc}

In Sect.~\ref{sec:anderson} through \ref{sec:lattice}, we will combine
the previous elements in circuits that simulate specific Hamiltonians.
In order to perform this task we must first introduce the mathematical
description of such circuits and the approximations under which they
will operate. We start considering the standard
Hamiltonian\cite{fazio01,bruder05} describing the physics of a
Josephson junction array such as the one in Fig.~\ref{fig:circuit}a
\begin{eqnarray}
  H &=& \sum_{ij,\sigma=\uparrow,\downarrow}\frac{1}{2}
   (n_{i\sigma} - n_{i\sigma}^{(g)})\bar{U}_{ij}(n_{j\sigma}-n_{j\sigma}^{(g)})
  \nonumber\\
  &+&  \sum_{ij,\sigma=\uparrow,\downarrow}
   \bar{J}_{ij} \cos(\phi_{i\sigma}-\phi_{j\sigma})
  + E(n_{0\uparrow},n_{0\downarrow}).
  \label{H-circuit}
\end{eqnarray}
Here, $n_{i\sigma}$ and $\phi_{i\sigma}$ are conjugate variables
denoting the excess number of Cooper pairs on an island and the
associated superconducting phase, respectively. $J_{ij}$ is the
Josephson energy for a junction that connects two islands. Finally,
there is a short range interaction term, $\bar{U}=4e^2C^{-1}$, which
is the inverse of the capacitance matrix. Similar to $\bar{J}$, $C$ is
a sparse matrix with nonzero elements on the diagonal, denoting the
intrinsic capacitance of each island, $C_{ii} \neq 0$, and also
between islands connected by junctions or capacitors. We choose all
capacitances to be small, $4e^2/C \gg \bar J$, and impose $C_{ii} \gg
C_{i\neq j}$ so that we can neglect any off-site interaction,
$\bar{U}_{i\neq j}\simeq 0.$

The residual charges, $n_i^{(g)}=n_i^{(0)}+n_i^{(DC)}$, are a
combination of a possible systematic shift, $n_i^{(0)},$ and an
effective term $n_i^{(DC)}=C_{i}^{(g)}V_{i}^{(g)}/2e$, proportional to
the potential applied to each island, $V_{i}^{(g)}$, and the
capacitance through which it is applied, $C_{i}^{(g)}$ [See
Ref.~\onlinecite{blais04}] From this it follows that the residual
charges can be set to zero by appropriate tuning the offset voltages
acting on each junction. As in current single and two-qubit
experiments \cite{vion02,wallraff04,yamamoto03,pashkin03,nakamura99},
we will work in the so called \textit{sweet spot} of energetic
degeneracy, given by $n^{(g)}_i=1/2$. Due to the low capacitance,
$\bar U\gg \bar J$, the state of each island is then constrained to
having 0 or 1 Cooper pair, or any linear superposition of these
states.

Contrary to the cavity-QED and superconducting qubit experiments
\cite{wallraff04} we would like the total number of Cooper pairs to be
a well defined quantum number. This means we do not want a residual
coupling between the states with 0 and 1 pairs, which would correspond
to a term $J_{ii}\cos(\phi_i)$ in the previous model. This is
effectively achieved by switching off the coupling between the
superconducting islands and the Cooper pair reservoir during the
experiment.

Finally, it is interesting to note that the total number of pairs in
the ground state can be tuned with very small voltages that take
islands away from degeneracy and which play the role of a local
chemical potential
\begin{equation}
  \mu_{i\sigma} = \sum_j \bar U_{ij} \left(n_{i\sigma}^{(g)}-\tfrac{1}{2}\right).
\end{equation}

\subsection{Hard-core boson model}
\label{sec:hard-core}

Summarizing the previous approximations, we have an array of
superconducting islands coupled both capacitively and via Josephson
tunneling. At the same time we have ensured that each superconducting
island has 0 o 1 excess Cooper pair, all other states being
energetically unfavorable. In other words, we have imposed the
hard-core bosons conditions on the Cooper pairs: they become like
impenetrable bosonic particles such that no two pairs with the same
effective ``spin''\footnote{That is, belonging to the same junction
  array.} can coexist on the same site.

In this limit we can describe any configuration of the circuit using
the occupation numbers $\{n_{i\sigma}\}.$ Due to the hard-core
condition, the dynamics is well approximated by the projection of the
circuit Hamiltonian (\ref{H-circuit}) onto the relevant energy
subspace, $n_{i\sigma}\in \{0,1\}.$ To perform this projection we will
need to express the phase terms in basis of occupation numbers
\begin{eqnarray}
  &&\sum_{ij\sigma}\bar J_{ij} \cos(\phi_{i\sigma}-\phi_{j\sigma}) =\\
  &&\quad\sum_{ij\sigma}\tfrac{1}{2}\bar J_{ij}
  \sum_{n_{i\sigma},n_{j\sigma}}|n_{i\sigma}+1,n_{j\sigma}-1\rangle
  \langle n_{i\sigma},n_{j\sigma}|
  + \mathrm{H.c.}\nonumber
\end{eqnarray}
Note that this equation has a very simple interpretation: Cooper pairs
are transferred one by one between any two islands connected by the
hopping matrix $\bar J_{ij}.$

The projection is implicitly performed by means of a more concise
mathematical representation that includes an exclusion principle for
particles on the same site. We introduce hard-core bosonic operators,
$a_{i\sigma}^\dagger$ and $a_{i\sigma}$, which respectively create and
annihilate a Cooper pair on a given site. They act on the basis
elements as follows
\begin{eqnarray}
  a_{i\sigma} |n_{i\sigma}=1\rangle = |n_{i\sigma}=0\rangle,\quad
  a_{i\sigma} |0\rangle = 0,\nonumber\\
  a_{i\sigma}^\dagger |0\rangle = |1\rangle,\quad
  a_{i\sigma}^\dagger |1\rangle = 0,\nonumber
\end{eqnarray}
have the usual bosonic commutation relations
\begin{equation}
  [a_{i\sigma}, a^\dagger_{j\sigma'}]=\delta_{ij} \delta_{\sigma \sigma'},\quad
  [a_{i\sigma}, a_{j\sigma'}]=0,
\end{equation}
and enforce the hard-core condition simply because
$a_{i\sigma}^{\dagger 2}=0.$ Using these operators we can express the
number of particles $n_{i\sigma} = a^{\dagger}_{i\sigma}a_{i\sigma}$
and the projected phase operator,
\begin{equation}
  2\cos(\phi_{i\sigma}-\phi_{j\sigma}) \simeq
  a^{\dagger}_{i\sigma} a_{j\sigma} + a^{\dagger}_{j\sigma} a_{i\sigma},
\end{equation}
which adopts the form of a hopping term.

\subsection{Single impurity Anderson model}
\label{sec:anderson}

We will apply all the approximations and techniques introduced
before to study a particular circuit, which consists of two impurity
islands coupled to two 1D Josephson junction arrays
[Fig.~\ref{fig:circuit}a]. The low energy dynamics of this circuit is
described by the bosonic Hamiltonian
\begin{eqnarray}
  H_A &=& -J\sum_{i\geq 1}
  (a^\dagger_{i+1\sigma} a_{i\sigma} + \mathrm{H.c}) - t \sum_\sigma
  (a^\dagger_{0\sigma}a_{1\sigma} + \mathrm{H. c.})
  \nonumber\\
  &+& E(n_{0\uparrow},n_{0\downarrow}) + \sum_{i\sigma} \mu_{i\sigma}
  n_{i\sigma}.
  \label{bosons}
\end{eqnarray}
Out of all terms in the hopping matrix $\bar J_{ij}$, we have left
only the Josephson energies between sites in the conduction array,
$\bar J_{ii+1}=J$ for $i>0$, and the coupling between the impurity
site and this band, $\bar J_{01}= \bar J_{10}=t$ [See
Fig.~\ref{fig:circuit}a]. Finally, as mentioned in
Sect.~\ref{sec:hard-core}, we have included an effective chemical
potential which depends on deviations from the degeneracy point of the
local potentials acting on each island.

We can map this Hamiltonian to a fermionic model using
another standard tool, the Jordan-Wigner transformation \cite{jordan28,mattis88}
\begin{eqnarray}
  \label{jw}
  a_{i\uparrow}  &=& c_{i\uparrow} (-1)^{\sum_{j<i} n_{j\uparrow}},\\
  a_{i\downarrow}  &=& c_{i\downarrow}
  (-1)^{\sum_{j<i} n_{j\downarrow} + N_{\uparrow}},\nonumber\\
  n_{i\sigma} &=& c^\dagger_{i\sigma}c_{i\sigma} =
  a^\dagger_{i\sigma}a_{i\sigma}.\nonumber
\end{eqnarray}
The new operators satisfy the usual fermionic anticommutation
relations $\{c_{i\sigma},c_{j\sigma'}^\dagger\} =
\delta_{ij}\delta_{\sigma\sigma'}$ and $N_{\sigma} = \sum_i
n_{i\sigma}.$ With this transformation, our effective Hamiltonian
(\ref{bosons}) converts \textit{identically} into a fermionic model
\begin{eqnarray}
  H_A &=& -J\sum_{i\geq 1}
  (c^\dagger_{i+1\sigma} c_{i\sigma} + \mathrm{H.c}) - t \sum_\sigma
  (c^\dagger_{0\sigma}c_{1\sigma} + \mathrm{H. c.})
  \nonumber\\
  &+& E(n_{0\uparrow},n_{0\downarrow}) + \sum_{i\sigma} \mu_{i\sigma}
  n_{i\sigma},
  \label{bosons}
\end{eqnarray}
which is known as the Anderson impurity model\cite{anderson61}. We
have thus demonstrated that the Anderson Hamiltonian accurately
describes the dynamics of a circuit that we have designed, under
reasonable approximations. Following the ideas sketched in the
introduction, we now argue that the quantum circuit can itself be used
to \textit{simulate} the dynamical and static properties of the
Anderson model. For instance, if we are able to create and cool this
circuit to its ground state, we will create a state which is related
to the ground state of the Anderson model by a unitary transformation.
The properties of the equivalent fermionic system can be recovered
from measurements of the Cooper pairs and the Jordan-Wigner relations
(\ref{jw}).

In the following subsections we will show how, by choosing other
limiting cases and designing slightly different circuits, one can
simulate not only the Anderson model, but other more complicated and
interesting Hamiltonians.

\subsection{Kondo model}

A special limit of the Anderson model is obtained when we impose a low
capacitance on the impurity, $U \to \infty$, and make the energy
displacement large compared to the coupling between the impurity and
its neighbors, $t \ll \epsilon$. Using standard second order
perturbation theory, the result is the single impurity Kondo
Hamiltonian \footnote{Note that this procedure, though equivalent in
  the results, differs in the method from \cite{schrieffer66}, where
  perturbation theory is carried on in momentum space.}
\begin{equation}
  H_K = -J \sum_{\langle i,j\rangle\sigma}  c^\dagger_{i\sigma} c_{j\sigma}
  + J_K \vec{S}_0 \cdot \vec{s}_{1}.
\end{equation}
Here $\vec{S}_0$ and $\vec{s}_{1}$ are the pseudo-spins of the impurity
and of the first lattice site
\begin{eqnarray}
  \vec{S}_0 =\sum_{\sigma,\sigma'}\frac{1}{2}\vec\sigma_{\alpha\sigma'}
  c^\dagger_{0\sigma}c_{0\sigma'},\; \mathrm{ and }\;
  \vec{s}_1 =\sum_{\sigma,\sigma'} \frac{1}{2}\vec\sigma_{\alpha\sigma'}
  c^\dagger_{1\sigma}c_{1\sigma'},
\end{eqnarray}
respectively, and $J_K =t^2/\epsilon$ is the anti-ferromagnetic
coupling between the Kondo impurity and the free fermions. We want to
stress that by tuning the parameters $J$, $t$ and $\epsilon$, we can
realize a crossover from a low energy regime below the Kondo
temperature $T_K$ to a high energy limit.

\begin{figure}
  \centering
  \includegraphics[width=0.85\linewidth]{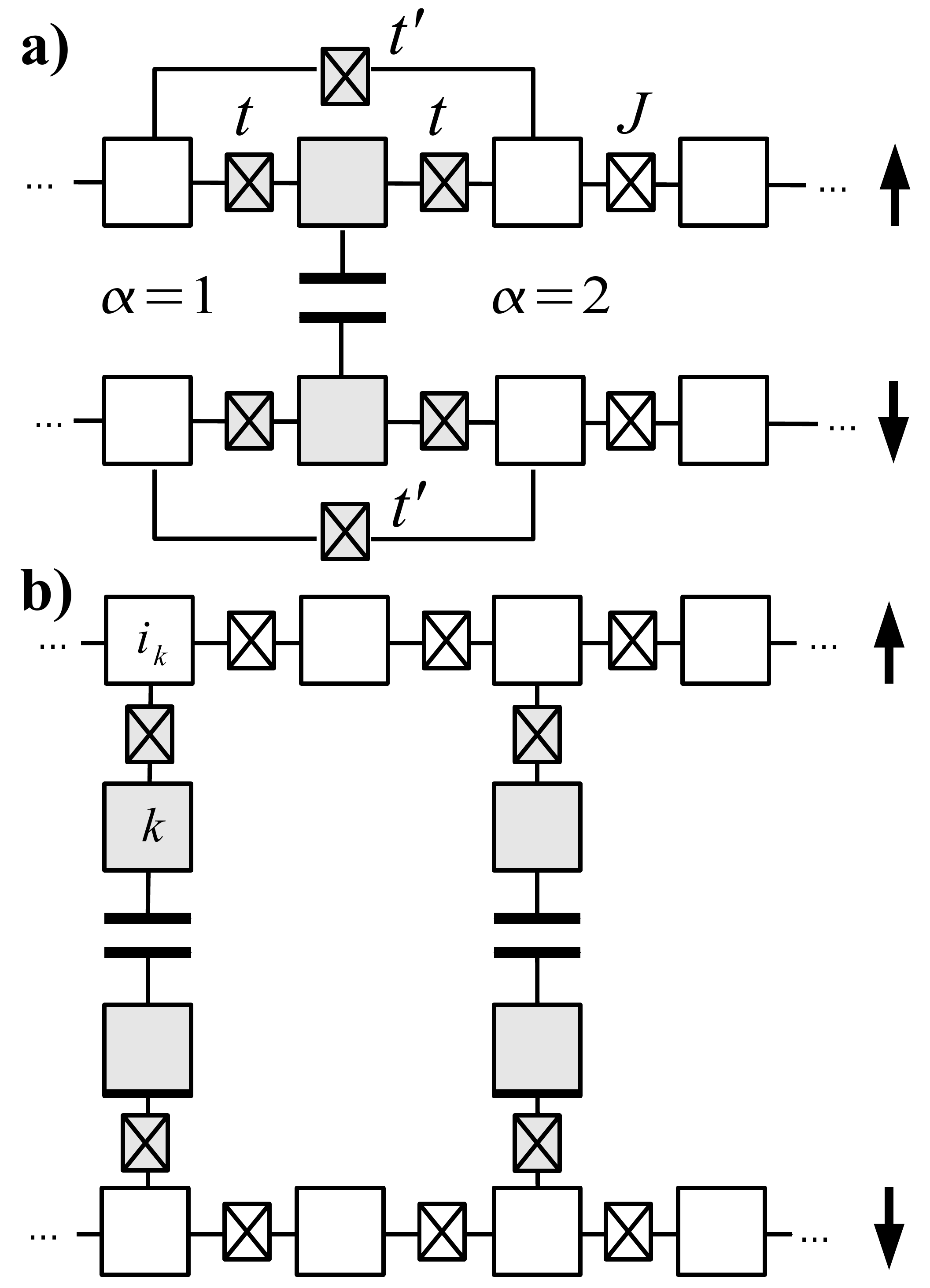}
  \caption{ (a) Circuit for a two-channel single impurity Kondo
    system. Each half of the circuit is associated to a different
    channel. An extra junction with negative sign, $t' \simeq -J_K$, is a
    quantum interference element that prevents hopping between
    channels. (b) Circuit for a 1D Kondo lattice.}
  \label{fig:2ck}
\end{figure}

\subsection{Two-channel Kondo model}

The previous setup and treatments can be reused to simulate a
two-channel Kondo system \cite{nozieres80}. As shown in
Fig.~\ref{fig:2ck}a, the impurity is in the middle of a
one-dimensional lattice, coupled to two different conduction bands.
Following similar steps, the new effective model becomes
\begin{equation}
  H_K = -J \sum_{\langle i,j\rangle\sigma,\alpha}  c^\dagger_{i\sigma\alpha} c_{j\sigma\alpha}
  + J_K \sum_{\alpha=1,2}\vec{S}_0 \cdot \vec{s}_{1\alpha},
\end{equation}
where $\alpha\in\{1,2\}$ denotes the two fermionic channels of our
problem. Note that in this Hamiltonian we have removed the terms that
induce hopping between different channels, $J_K
c^\dagger_{1\sigma\alpha'}c_{1\sigma\alpha}$. These and higher order
couplings are canceled by two cleverly placed small junctions whose
tunneling has been magnetically calibrated to the value
$t'\simeq-J_K$.

\subsection{Kondo lattice}
\label{sec:lattice}

A slightly less trivial circuit is required to simulate a Kondo
lattice that has multiple localized impurities interacting with
fermions. There are two different ways of doing it, leading to
slightly different physics in each case. A simple concatenation of the
circuit in Fig.~\ref{fig:2ck}a will produces a set of impurities
coupled by a number of sites with free fermions, equivalent to the
model introduced by Paredes\cite{paredes05} for cold atoms. However,
the Kondo lattice as known in the literature \cite{tsunetsugu97}
considers multiple impurities coupled to a \textit{common} conduction
band, which can be achieved using the circuit in Fig.~\ref{fig:2ck}b.
Following Sect.~\ref{sec:hard-core} we can write down an effective
bosonic Hamiltonian \begin{eqnarray}
  H &=& -J\sum_{\langle i,j\rangle} a^\dagger_{i\sigma} a_{j\sigma}
  - t \sum_k (b^\dagger_{k\sigma} a_{i_k\sigma} + \mathrm{H.c})\nonumber\\
  &+& \sum_k E(n^{(b)}_{k\uparrow},n^{(b)}_{k\downarrow}) +
  \sum_{i\sigma} \mu_{i\sigma} n^{(a)}_{i\sigma},
\end{eqnarray}
where we have introduced different hard-core bosonic operators for
pairs in the conduction band, denoted by
$a_{i\sigma},a^\dagger_{i\sigma},$ and $n^{(a)}_{i\sigma} =
a_{i\sigma}^\dagger a_{i\sigma}$, and for pairs in the impurities,
given by $b_{i\sigma},b^\dagger_{i\sigma},$ and $n^{(b)}_{k\sigma} =
b_{k\sigma}^\dagger b_{k\sigma}$.

This circuit does not have a direct translation into a fermionic
Anderson model because for such a quasi-2D structure there is no
Jordan-Wigner transformation that preserves the shape of the bosonic
Hamiltonian. We will however perform an incomplete Jordan-Wigner
transformation which acts only on the conduction arrays,
$a_{i\sigma}$, leaving the $b_{k\sigma}$ operators untouched,
\begin{eqnarray}
  a_{i\uparrow} &=& c_{i\uparrow} (-1)^{\sum_{j<k} n^{(a)}_{j\uparrow}},\\
  a_{i\downarrow} &=& c_{i\downarrow} (-1)^{\sum_{j<k} n^{(a)}_{j\downarrow}+ N^{(a)}_\uparrow},\nonumber
\end{eqnarray}
where $N^{(a)}_\sigma = \sum_i n_{i\sigma}^{(a)}.$ With this unitary
map, and working out the second order perturbation theory, one arrives
to the standard Kondo lattice model
\begin{equation}
  H_{KL} = -J \sum_{\langle i,j\rangle,\sigma} c^\dagger_{i\sigma}c_{j\sigma}
  + \sum_{k,\sigma,\sigma'} \frac{J_K}{2}\vec{S}_k\cdot \vec\sigma_{\alpha\sigma'}
  c^\dagger_{1\sigma}c_{1\sigma'} .
\end{equation}
Here, $\vec{S}_k$ denotes the spin of the $k$-th impurity which is
coupled to the site $i_k$ and is defined using bosonic operators $
\vec{S}_{i_k}
=\sum_{\sigma,\sigma'}\frac{1}{2}\vec\sigma_{\alpha\sigma'}
b^\dagger_{i_k\sigma}b_{i_k\sigma'}. $ This general notation allows
for having less impurities than conduction electron, as well as them
being placed on arbitrary positions. The coupling is once more
anti-ferromagnetic $J_K \sim t^2/\epsilon$ and just as tunable as in
the case of a single impurity.

\section{Energy scales and decoherence}

Let us consider what are the experimental challenges for implementing
these mappings. We begin with the energy scales of our problem. On the
one hand we have free fermions, which hop with a tunneling amplitude
$J$ and a Fermi energy $\epsilon_F\sim 2J$. In current experiments
with charge qubits $J$ can be roughly of order of a few gigahertz. On
the other hand we have the Kondo coupling $J_K = t^2/\epsilon$, which
will be in general smaller. For this coupling to be relevant, the
temperature at which the experiment is performed should lay below the
Kondo Temperature, $k_B T_K= \epsilon_F \exp[1/J_K\rho(k_F)]$, where
$\rho(k_f)$ is the density of states at the Fermi energy. Since in our
case the density of states at half filling ($\mu=0$) is roughly
$\rho(k_f) = \pi/J$, using $t\sim J$, we have $k_B T_K \sim 2J
\exp(-\epsilon/J\pi)$. This means that, conservatively, for
$\epsilon/J = 8-10$, the Kondo temperature $T_K \sim 40-75$mK would be
larger $20$ mK, the refrigerator temperature of current experiments
\cite{wallraff04}.

Another important challenge is decoherence, which in our setup arises
mainly from charge fluctuations. As is the case of scalable quantum
computation, the characterization of decoherence in multi-qubit setups
is still an open problem that deserves further investigation. As a
guide, we consider the time-scale for the decoherence of a single
qubit is $T_2 \sim 0.5\mu s$ \cite{vion02,wallraff05}, which is
achieved for dispersive readouts. This time-scale gives a frequency of
2 MHz which is well below $J/h\sim 10$GHz and two orders of magnitude
away of the lowest frequency $J_K/h=t^2/\epsilon\sim 100$ MHz that we
find here. Furthermore, associated to this decoherence rate we can
establish an effective temperature of $0.1$ mK which is also well
below the Kondo temperature.

When compared to other systems, such as cold atoms in optical lattices
\cite{jaksch98}, the superconducting circuit approach looks as
enjoying real advantages. First, the superconducting qubits are not
constrained to periodic or quasi-periodic structures. Second and most
important, the introduced setup has the potential quality of a good
quantum simulator, which is measured by the number of times a particle
can hop before the wave function is affected by decoherence. For cold
atoms in the lowest band of an optical lattice \cite{greiner02},
coherence times of 500 ms are to be compared with a hopping rate of
1kHz between lattice sites, giving a ratio of 500 or better. A similar
ratio with superconducting circuits, where the hopping is around 10
GHz, means the decoherence time can be 0.05$\mu$s, an order of
magnitude faster than current experiments \cite{vion02,wallraff05}. On
the other hand, superconducting circuits have the disadvantage of
being one-time experiments because the geometry of the couplings
cannot be modified in real-time. In addition, it may be complicated to
calibrate all islands to reach the sweet spot. However, if we realize
that the couplings can be initially switched to zero, this task is as
difficult as tuning the individual superconducting qubits in a
scalable quantum computer.

\section{Measurement}

While we have demonstrated that certain quantum circuits can be used
to simulate interesting fermionic models, the equivalence between
circuit and the Hamiltonian involves the use of Jordan-Wigner
transformations (\ref{jw}). Therefore, not all measurements on the
quantum circuit will give directly properties of the respective
fermionic system. One can envision three kinds of measurements in
these systems which do not have this problem. The first one are global
transport measurements that mimic past experiments with Josephson
junction arrays \cite{oudenaarden96}. Since they are based on total
particle numbers, they are the same for fermionic and bosonic models.

Next we have spectroscopic measurements and measurement of energy
gaps. The coupling between the impurity and the conduction band leads
to a binding energy of order $\Delta = k_BT_K$ given above. This
energy can be measured by trying to polarize the impurity with an
oscillating voltage applied to the island. We expect this to have an
effect either when the potential is of order $\Delta$ or when it
oscillates with a frequency of order $\Delta/\hbar$.

The third kind of experiments corresponds to measuring the individual
qubits and extracting information about certain correlation functions.
In the case of Fig.~\ref{fig:circuit}a, the formation of a singlet
with the conduction band electrons gives rise to a strong spin
correlation in the vicinity of the impurity. The correlation length is
of order $\xi =\exp[1/J_K\rho(k_F)]$ sites and can be seen in the
connected correlator as a function of the site $m$ \begin{equation}
  C_m = \langle S_{0z}s_{mz}\rangle - \langle S_{0z}\rangle\langle
  s_{mz}\rangle \sim \exp(-m/\xi).
\end{equation}
This is known as the Kondo cloud and has not been directly detected by
condensed matter setups. In our case, $C_m$ can be computed from the
statistics of the population difference between islands, $s_{iz} =
(n_{i\uparrow}-n_{i\uparrow})/2$, which is best measured using
dispersive probes~\cite{johansson06,wallraff05}.

\section{Conclusions}

Summing up, in this paper we have presented a mapping between some
superconducting circuits to a family of relevant Hamiltonians
describing magnetic impurities in conduction bands. While there are
still implementation issues to be solved, they seem less demanding
than those of a fully scalable quantum computer implementation with
the same technology. We thus expect that our work will motivate
further developments, both on the experimental and on the theoretical
side, searching better multi-qubit coherence times and other mappings
which are based on flux or hybrid qubits.

We thank Vitaly Golovach, Stefan Kehrein, and Jan von Delft for their
comments. E.S. acknowledges financial support from DFG SFB 631, EU
EuroSQIP projects, and the German Excellence Initiative via the
``Nanosystems Initiative Munich (NIM)''. M.A.M.D. and
J.J.G.R. acknowledge financial support from projects FIS2006-04885
(Spanish M.E.C.) and CAM-UCM/910758. J.J.G.R. acknowledges support
from the Ramon y Cajal Program of the M.E.C.

\end{document}